\documentclass[
6x9    ,final            
  ]
  {aipproc}

\layoutstyle{6x9}

\begin{document}

\title{Positrons at Jefferson Laboratory}


\author{Anthony W. Thomas}{
  address={Suite 1, Jefferson Lab,12000 Jefferson Ave., Newport News VA 23606 USA \\ and College of William and Mary, Williamsburg VA 23187 USA}
}

\begin{abstract}
We review the compelling case for establishing a capability to accelerate 
positrons at Jefferson Lab. 
The potential appplications range from the study of two-photon 
exchange and deeply-virtual Compton 
scattering to exploiting the charge current weak 
interaction to probe the flavor structure of hadrons and 
nuclei. There are also fascinating ideas for using 
such a capability to discover new physics beyond the 
Standard Model of nuclear and particle physics.
\end{abstract}

\maketitle


\section{Introduction}

While there are really only a few reasons for tackling the 
difficult and expensive challenge of 
providing positron beams at a laboratory like Jefferson Lab, 
they are compelling. In important circumstances one can use the 
interference with single photon exchange to clearly isolate 
a particular piece of physics. Examples include 
the study of the two-photon exchange contribution to 
elastic scattering and the extraction of information 
on the generalized parton distributions (GPDs). The charge current processes, 
$(e^-,\nu_e)$ and $(e^+,\bar{\nu}_e)$, 
uniquely isolate either positively or negatively charged quarks, 
respectively, and can therefore provide important information on the flavor structure of nucleons and 
nuclei. Finally,  there are a number of interesting possibilities, including $e^+ - e^-$ annihilation, which have the 
potential to reveal physics outside of that described within the Standard Model. We set the stage for the main 
presentations at the workshop by outlining just a few of the possible highlights.

\section{Interference}

One of the technical triumphs of the JLab program, based on its high duty factor, high polarization 
and intensity, is that one can measure recoil polarization in processes like $(e,e^\prime N)$. The application of 
this process to elastic $e-p$ scattering led to the discovery that, contrary to the conclusions based on 
using a Rosenbluth separation, the shape of the electric and magnetic form factors of the proton was very 
different. Indeed, the ratio $G_E/G_M$ drops rapidly towards zero as $Q^2$ increases 
towards 7 - 8 GeV$^2$~\cite{Punjabi:2005wq}. 
Extensive theoretical studies have shown that this difference is most likely a 
consequence of two-photon exchange, which seems to have a much bigger effect on the Rosenbluth separation 
than on recoil polarization~\cite{Blunden:2005ew,Arrington:2007ux}. 
Experimental confirmation of this explanation would be simple if one could compare 
$e^+$ and $e^-$ scattering, with the sign of the interference term being different in these two cases. 

Turning to a different topic, the challenge of 
resolving the so-called ``spin crisis'' has been with us for more than 
20 years~\cite{Ashman:1987hv}. 
Those two decades of experimental work have provided an 
enormous amount  of information and in 
many ways the crisis has now been 
resolved~\cite{Thomas:2008bd,Myhrer:2007cf}. From 
accurate measurements of the proton spin 
structure function, $g_1^p(x)$, we now know that 
the quarks carry about one third of the proton spin, not zero 
as appeared possible back in 1988. 
We also know that the spin carried by polarized gluons in the proton is 
an order of magnitude smaller than that required 
to resolve the crisis through the $U_A(1)$ axial 
anomaly. On the other hand, crucial 
non-perturbative aspects of hadron structure, namely relativistic quark 
motion, the hyperfine interaction in QCD~\cite{Myhrer:1988ap} 
and the pion cloud required by chiral 
symmetry~\cite{Schreiber:1988uw}, all act to reduce 
the spin carried by the quarks. These 
effects account fully for the modern proton spin sum and hence resolve 
the spin crisis.

Of course, one really wants to confirm this theoretical explanation by independent experimental tests. Key to 
this is the realization that all of the mechanisms just outlined have the effect of replacing quark spin by 
quark and anti-quark orbital angular momentum~\cite{Thomas:2008ga} 
and the ideal method to access that experimentally is 
using the GPDs~\cite{Ji:1995cu}. 
The GPDs are extracted from deeply-virtual 
Compton scattering (DVCS) through the 
interference between the Bethe-Heitler process and DVCS~\cite{Roche:2000ng}. 
This interference changes sign when one switches 
from $e^-$ to $e^+$ and this is critical to being sure that one has correctly recognized the interference term 
rather than the square of the DVCS amplitude. 
With its unique access to the critical valence region JLab, with 
its upgrade to 12 GeV, is ideally placed to determine the amount of orbital angular momentum carried by the 
quarks and access to high intensity beams of 
polarized positrons would be extremely valuable.

\subsection{Charged current measurements}

Another major technical development at JLab has been 
the ability to perform remarkably precise 
measurements of parity violating electron scattering (PVES). 
This has enabled the accurate determination 
of the strange vector current matrix elements in the 
proton~\cite{Armstrong:2005hs,Aniol:2005zf,Maas:2004dh}, 
with the result that strange quarks contribute 
less than 5\% of the magnetic moment or charge radius squared of the 
proton~\cite{Young:2006jc,Young:2007zs}. Such measurements are the 
QCD equivalent of the determination of the Lamb shift in QED, as the strange quarks are not part of the 
valence structure of the proton and contribute only through vacuum fluctations. Calculations based on lattice QCD 
and modern chiral extrapolation techniques are in excellent agreement with the experimental results, albeit 
with errors an order of magnitude smaller~\cite{Leinweber:2004tc,Leinweber:2006ug} -- a 
unique occurence in modern strong interaction physics.

In contrast with the strange elastic form factors, the experimental determination of the difference in the 
$s$ and $\bar{s}$ parton distributions have proven far more difficult~\cite{Mason:2007zz}, 
largely as a consequence of the 
difficulties associated with neutrino experiments. Such a difference was anticipated more than 20 years ago 
on the basis of the fluctuation $p \rightarrow K^+ \Lambda$~\cite{Signal:1987gz}. 
Indeed, it is now known that the chiral 
non-analytic behavior of $s(x)$ and $\bar{s}(x)$ is different and hence, $s(x) - \bar{s}(x)$ must be  
non-zero~\cite{Thomas:2000ny}. High luminosity electron and positron beams, preferably 
at a future electron-hadron collider such as ELIC proposed at JLab, would permit one to accurately map 
this difference, which also has consequences for the search for new physics in phenomena such as the 
NuTeV anomaly~\cite{Zeller:2001hh}.

The quest to understand the role of the quark and gluon degrees of freedom in defining the properties of 
atomic nuclei is one of the great challenges facing modern nuclear physics. The nuclear EMC effect, which was 
discovered more than a quarter of a century ago~\cite{Aubert:1983xm}, 
provides compelling evidence that the structure of a 
bound ``nucleon'' differs in a fundamental way from its free counterpart~\cite{Geesaman:1995yd} 
and yet this evidence is largely ignored 
as inconvenient by the community. The quark-meson coupling (QMC) model is an extremely efficient and 
effective formalism with which to investigate the role of quarks and gluons in determining nuclear 
properties~\cite{Guichon:1987jp,Saito:2005rv}, from the saturation of nuclear 
matter to the EMC effect.  Within the QMC model, and its modern 
incarnation based upon the chiral symmetric, covariant, confining NJL model, developed by Bentz, Clo\"et and 
Thomas~\cite{Bentz:2001vc}, the paradigm which has underpinned 
nuclear theory for over 70 years is replaced by the understanding 
that what occupies shell model orbits in atomic nuclei are not free neutrons and protons but quark clusters 
with nucleon quantum numbers whose structure has self-consistently adjusted to the local mean scalar and 
vector fields in the nuclear medium~\cite{Guichon:1995ue}. 
This picture has been successfully linked to conventional many-body theory 
through the derivation of an equivalent energy functional which yields a remarkably successful 
density-dependent force of the Skyrme type. Recently applications of the QMC model include an 
explanation of the experimental absence of $\Sigma$-hypernuclei~\cite{Guichon:2008zz,Hashimoto:2006aw}, 
the prediction of weak binding for 
$\Xi$-hypernuclei, interesting predictions for the photo-production of 
hypernuclei~\cite{Tsushima:2009zh} and new results for the 
properties of dense nuclear matter in $\beta$-equilibrium, including hyperons~\cite{RikovskaStone:2006ta}.

In the present context, it is especially interesting to note the discovery of the isovector EMC effect by 
Clo\"et {\it et al.}~\cite{Cloet:2009qs}. 
In particular, even if one makes an ``iso-scalarity'' correction 
to the structure function of Fe by subtracting the 
structure functions of the 4 extra neutrons, one cannot eliminate so easily their effect on all of the 
remaining neutrons and protons. Because of the isovector repulsion between $d$-quarks and the attrraction 
between $d$'s and $u$'s, the $d$-quarks will have a different distribution in Bjorken-$x$ than the 
$u$-quarks. The sign of this effect is obviously the same as that associated with normal charge symmetry 
violation (CSV)~\cite{Londergan:2003ij,Rodionov:1994cg}, 
itself an important object of study in modern hadron physics, and together these two effects 
(the ``pseudo-CSV'' associated with the iso-vector EMC effect and regular CSV associated with $u-d$ mass 
differences), can account for the NuTeV anomaly. Far from being a disappointment, in the sense that there 
is no evidence for physics beyind the Standard Model, this support for an iso-vector EMC effect provides 
powerful support for this new paradigm for nuclear structure, itself a remarkably important issue.

The unambiguous confirmation of  the difference predicted by Clo\"et {\it et al.} between $d_A$ and $u_A$ 
in a nucleus with N $\neq$ Z, even after correcting for the excess neutrons (or protons), is possible using 
a comparison of charge current deep-inelastic scattering. Once again, a high energy, high luminosity 
collider such as that proposed at JLab, would be ideal. Other tests which have been proposed include 
the comparison of semi-inclusive $\pi^+$ and $\pi^-$ 
deep-inelastic scattering at 12 GeV at JLab~\cite{JLab} but the 
complications of final state interactions have the potential to complicate the interpretation there. This is not 
an issue for the charged current comparison.

\section{Beyond the Standard Model}

A comparison of the charged current cross section for electrons and positrons as function of polarization 
at HERA showed the potential for testing the Standard Model in this way~\cite{Zhang:2007pp}. 
However, lack of luminosity 
meant that it was never competitive with other methods. An electron-ion collider with luminosity at the 
level of $10^{35}$ cm$^2$/sec, as proposed at JLab, may well turn this around.

Another rather different approach will be described at this meeting by 
Bogdan Wojtsekhowski~\cite{Bogdan}. There have 
been suggestions for some time that an excess of 511 KeV X-rays from the galactic center may signal the 
existence of a new, light $U$-boson with mass in the range of a few to tens of MeV. The most promising 
method to search for such a particle would be in $e^+ - e^-$ annihilation using an intense positron beam, 
at JLab possibly at the FEL. For more details we refer to the presentation of Wojtsekhowski. 

\section{Conclusion}

More details of the ideas presented here and a number of others will be found in these proceedings. 
Even from this introductory taste, it should be clear that the initiative to develop a positron capability at 
JLab is very well motivated scientifically. There are some very important issues that can be addressed and 
resolved only in this way and I am sure that these will prove compelling when it comes to finding the 
necessary resources.


\begin{theacknowledgments}
This work was supported by U.S. DOE Contract No. DE-AC05-06OR23177, 
under which Jefferson Science Associates, LLC operates Jefferson 
Laboratory.
\end{theacknowledgments}


\end{document}